\documentclass[a4paper]{jpconf}
\usepackage{graphicx}
\usepackage{amsmath}
\usepackage{color}
 \usepackage{amsfonts}
 \usepackage{bm}
\usepackage{epsfig,color,graphicx}
\usepackage{amssymb}
\usepackage{array}
\usepackage{booktabs} 
\newcommand{\be}{\begin{eqnarray}}
\newcommand{\ee}{\end{eqnarray}}
\newcommand{\bee}{\begin{eqnarray*}}
\newcommand{\eee}{\end{eqnarray*}}

\def\pa{\partial}

\let\na=\nabla
\let\pa=\partial

\let\gd=\delta
\let\gD=\Delta
\let\ge=\epsilon
\let\gf=\varphi
\let\gh=\eta
\let\gj=\phi
\let\gm=\mu
\let\gp=\pi
\let\gq=\theta
\let\gs=\sigma
\let\gt=\tau
\let\gw=\omega
\let\gy=\psi
\let\gz=\zeta

\let\na=\nabla
\let\pa=\partial

\let\gd=\delta
\let\gD=\Delta
\let\ge=\epsilon
\let\gf=\varphi
\let\gh=\eta
\let\gj=\phi
\let\gl=\lambda
\let\gm=\mu
\let\gp=\pi
\let\gq=\theta
\let\gr=\rho
\let\gs=\sigma
\let\gt=\tau
\let\gw=\omega
\let\gW=\Omega
\let\gy=\psi
\let\gx=\xi
\let\gz=\zeta

\let\lbq=\label

\let\na=\nabla

\let\lbq=\label

\let\na=\nabla

\let\gG=\Gamma

\def\wsi{{\bf i}}
\def\wsj{{\bf j}}
\def\wsn{{\bf n}}

\def\vslc{\vec{\mathcal{\ell}}}

\def\waA{{\bf A}}
\def\waB{{\bf B}}

\def\R#1#2{\frac{#1}{#2}}
\def\btbl{\begin{tabular}}
\def\etbl{\end{tabular}}
\def\bqbl{\begin{eqnarray}}
\def\eqbl{\end{eqnarray}}

\def\thesection {\arabic{section}}
\def\thesubsection {\thesection.\arabic{subsection}}
\def\thesubsubsection {\thesubsection.\arabic{subsubsection}}

\def\begindoc{\begin{document}}
\def\enddoc{\end{document}}
\def\bskip{\baselineskip}
\def\NN{\nonumber\\}
\def\tensor#1{\stackrel{\Leftrightarrow}{#1}}
\def\bq{\begin{equation}}
\def\eq{\end{equation}}
\def\bc{\begin{center}}
\def\ec{\end{center}}
\def\beginbib#1{\begin{thebibliography}{#1}}
\def\endbib{\end{thebibliography}}
\def\rfb#1({Ref.\cite{#1})}
\def\dd#1{\frac{d}{d#1}}
\def\dfr#1#2{\frac{#1}{#2}}
\def\bqy{\begin{eqnarray}}
\def\eqy{\end{eqnarray}}
\def\beginenum{\begin{enumerate}}
\def\endenum{\end{enumerate}}
\def\negl#1{\left\{#1\right\}_\Downarrow}
\def\ave#1{\left<#1\right>}

\mathchardef\muchg="321D

\let\ga=\alpha
\let\gb=\beta
\let\gc=\chi
\let\gd=\delta
\let\ge=\epsilon
\let\gf=\varphi
\let\gg=\gamma
\let\gh=\eta
\let\gi=\iota
\let\gj=\phi
\let\gk=\kappa
\let\gl=\lambda
\let\gm=\mu
\let\gn=\nu
\let\go=o
\let\gp=\pi
\let\gq=\theta
\let\gr=\rho
\let\gs=\sigma
\let\gt=\tau
\let\gu=\upsilon
\let\gv=\varpi
\let\gw=\omega
\let\gx=\xi
\let\gy=\psi
\let\gz=\zeta
\let\gA=A
\let\gB=B
\let\gC=X
\let\gD=\Delta
\let\gE=E
\let\gF=\Phi
\let\gG=\Gamma
\let\gH=H
\let\gI=I
\let\gJ=\vartheta
\let\gK=K
\let\gL=\Lambda
\let\gM=M
\let\gN=N
\let\gO=O
\let\gP=\Pi
\let\gQ=\Theta
\let\gR=R
\let\gS=\Sigma
\let\gT=T
\let\gU=\Upsilon
\let\gV=\varsigma
\let\gW=\Omega
\let\gX=\Xi
\let\gY=\Psi
\let\gZ=Z

\def\dga{{\dot{\alpha}}}
\def\dgb{{\dot{\beta}}}
\def\dgc{{\dot{\chi}}}
\def\dgd{{\dot{\delta}}}
\def\dge{{\dot{\epsilon}}}
\def\dgf{{\dot{\varphi}}}
\def\dgg{{\dot{\gamma}}}
\def\dgh{{\dot{\eta}}}
\def\dgi{{\dot{\iota}}}
\def\dgj{{\dot{\phi}}}
\def\dgk{{\dot{\kappa}}}
\def\dgl{{\dot{\lambda}}}
\def\dgm{{\dot{\mu}}}
\def\dgn{{\dot{\nu}}}
\def\dgo{{\dot{o}}}
\def\dgp{{\dot{\pi}}}
\def\dgq{{\dot{\theta}}}
\def\dgr{{\dot{\rho}}}
\def\dgs{{\dot{\sigma}}}
\def\dgt{{\dot{\tau}}}
\def\dgu{{\dot{\upsilon}}}
\def\dgv{{\dot{\varpi}}}
\def\dgw{{\dot{\omega}}}
\def\dgx{{\dot{\xi}}}
\def\dgy{{\dot{\psi}}}
\def\dgz{{\dot{\zeta}}}
\def\dgA{{\dot{A}}}
\def\dgB{{\dot{B}}}
\def\dgC{{\dot{X}}}
\def\dgD{{\dot{\Delta}}}
\def\dgE{{\dot{E}}}
\def\dgF{{\dot{\Phi}}}
\def\dgG{{\dot{\Gamma}}}
\def\dgH{{\dot{H}}}
\def\dgI{{\dot{I}}}
\def\dgJ{{\dot{vartheta}}}
\def\dgK{{\dot{K}}}
\def\dgL{{\dot{\Lambda}}}
\def\dgM{{\dot{M}}}
\def\dgN{{\dot{N}}}
\def\dgO{{\dot{O}}}
\def\dgP{{\dot{\Pi}}}
\def\dgQ{{\dot{\Theta}}}
\def\dgR{{\dot{R}}}
\def\dgS{{\dot{\Sigma}}}
\def\dgT{{\dot{T}}}
\def\dgU{{\dot{\Upsilon}}}
\def\dgV{{\dot{\varsigma}}}
\def\dgW{{\dot{\Omega}}}
\def\dgX{{\dot{\Xi}}}
\def\dgY{{\dot{\Psi}}}
\def\dgZ{{\dot{Z}}}

\newfont{\wncyrIII}{wncyr8 scaled\magstep1}
\newfont{\wncybIII}{wncyb8 scaled\magstep1}
\begin{document}
\title{Simulation of the electromagnetic wall response during Vertical Displacement Events (VDE) in ITER tokamak}

\author{C\v{a}lin V. Atanasiu$^1$, Leonid E. Zakharov$^2$, Karl Lackner$^3$, Matthias Hoelzl$^3$}

\address{$^1$National Institute for Laser, Plasma and Radiation Physics,  Atomistilor 409, P.O. Box MG-36, 077125 Magurele-Bucharest, Romania }

\address{$^2$LiWFusion, P.O. Box 2391, Princeton,  NJ 08543, USA}

\address{$^3$Max Planck Institute for Plasma Physics,  Boltzmannstr. 2, 85748 Garching, Germany}

\ead{$^1$cva@ipp.mpg.de}
\begin{abstract}
The key basis for tokamak plasma disruption modeling is to understand how currents flow to the plasma facing surfaces during plasma disruption events. In ITER tokamak, the occurrence of a limited number of major disruptions will definitively damage the chamber with no possibility to restore the device. In the current exchange plasma-wall-plasma, according to the Helmholtz decomposition theorem, our surface current density in the conducting shell - the unknown of our problem - being a vector field twice continuously differentiable in 3D, has been splited into two components: an irrotational (curl-free) vector field and a solenoidal (divergence-free) vector field. Developing a weak formulation form and  minimizing the correspondent energy functionals in a Finite Element approach, we have obtained the space and time distribution of the surface currents. We verified successfully our numerical simulation with an analytical solution with pure homogeneous Neumann B.C. and satisfying the necessary existence condition. By considering the iron core presence in JET tokamak,  we have split the magnetization currents - the unknowns in some integral equations - into two components, the first producing a magnetic field in the iron region only and the second producing a magnetic field in the vacuum, obtaining thus a better evaluation of the influence of the iron core on the plasma equilibrium. To reduce the influence of the singularities appearing during the surface currents determination in multiply connected domains (L-shaped domains) we have used a conformal transformation method. 
\end{abstract}

\section{Introduction}

Plasma disruptions in tokamaks represent a significant obstacle in enhancing performance of the plasma regime. In ITER tokamak, the occurrence of a limited number of major disruptions will definitively damage the chamber with no possibility to restore the device. In the next step machines, such as ITER, disruptions impose very challenging requirements on the design of the structural elements of the machine and its in-vessel components. Therefore, empirical 
approaches for determining the operational space for high performance and, at the same time, disruption-free regimes are excluded. Theoretical and modelling approaches by using the present level of experiments are necessary \cite{LEZ2008, LEZ2012, MAT1,MAT2}. 

It is well known that the necessarily large toroidal currents in tokamak concept suffers from a fundamental problem of stability. The Wall Touching Kink Mode (WTKM) - a nonlinear MHD instability - leads to a dramatic quench of the plasma current within $ms$:  very energetic electrons are created (runaway electrons) and finally a global loss of confinement happens, i.e. a major disruption. The WTKM are frequently excited during the Vertical Displacement Event (VDE) and cause big sideways forces on the vacuum vessel \cite{RIC2009, RIC2000}.

Understanding that in disruptions the sharing of electric current between the plasma and the wall plays an important role in plasma dynamics \cite{LEZ2008, LEZ2012, LEZ20131, LEZ20132}, we have developed  a wall model that covers both eddy currents, excited inductively, and source/sink currents due to current sharing between the plasma and the wall.
We have adopted a triangle representation of the plasma facing wall surface \cite{LEZ2015} (simplicity and analytical formulas for magnetic field {\bf B} and magnetic vector potential {\bf A} of a uniform current in a single triangle). We have considered the wall in its thin wall approximation
(reasonable for thin stainless steel structures of the vacuum vessel of about 1-3 cm thick with conductivity  $\sigma=1.38·10^6\; \Omega^{-1}m^{-1}$).  

In section 2, we define our electromagnetic thin-wall model describing both surface current components. Section 3 describes the weak formulation (under a finite element frame) to determine the unknowns of our problem - the surface currents in the wall. In section 4, numerical and analytical examples are 
presented. The influence of the presence of an iron core transformer tokamak (like in JET tokamak) is described in section 5. The summary is given in section 6.
\section{Electromagnetic thin-wall model}
According to   Helmholtz decomposition theorem, our surface current density 
$d_w{\bf j}$ in the conducting shell (a vector field twice continuously differentiable in 3D) can be split into two components: an irrotational (curl-free) vector field and a solenoidal (divergence-free) vector field \cite{LEZ2015} 
\be 
\begin{aligned}
d_w{\bf j}={\bf i}-d_w\sigma\nabla \phi^S, \\ \\
{\bf i}=\nabla I\times {\bf n},\;\; (\nabla \cdot {\bf i}=0),
\end{aligned}
\ee
where $ {\bf i} $ is  the divergence free surface current (eddy currents), 
$−d_w\sigma\nabla \phi^S$ is the source/sink current (S/SC) with potentially finite divergence in order to describe the current sharing between plasma and wall, $\sigma$ is surface wall conductivity, $d_w$ represents the thickness of the current distribution, $I$ is the stream function of the divergence free component (eddy currents), ${\bf n}$ is the unit normal vector to the wall while $\phi^S$ is the source/sink potential (a surface function).  

The S/S-current in Eqs.\;(1) is  determined from the continuity equation of the S/S
 currents across the wall 
\be
\na\cdot (d_w\wsj)
=
-\na\cdot(d_w\sigma\na\gj^S)
=
j_\perp,
\lbq{eq:gd}
\ee
   where $j_\perp
\equiv
-(\wsj\cdot\wsn)$ is the density of the current coming from/to
the plasma, 
$j_\perp > 0$ for $ j_{\perp}$ flowing from the plasma to the wall. 

Faraday law gives
\be
-\R{\pa\waA}{\pa t} -\na\gj^E
=
\frac{1}{d_w\sigma}(\na I\times\wsn)-\na\gj^S
, 
\ee
where $\waA$ is the magnetic vector potential of the magnetic field {\bf B}, while $\gj^E$ is the electric potential. 
Equations\,(2, 3) describe the current distribution
in the thin wall given the following sources: the current density coming from/to
the plasma $j_\perp$, the normal to the wall components of the magnetic field due to the plasma  
($B^{pl}_\perp$), to different coils ($B^{coil}_\perp$) and to the iron core transformer ($B_{\perp}^{Fe}$) all as functions of position and time. It is to note 
 that  Equation\,(2) for $\gj^S$ is independent from Equation\,(3), but contributes via 
 $\pa B^S_\perp/\pa t$ to the r.h.s. of Equation\,(3).
 
 In our finite element solving approach, with  a uniform 
surface current $\wsj$=const inside each triangle, the magnetic vector potential has been calculated with the relation
 \be
{\bf A}^{wall}({\bf r})={\bf A}^{I}({\bf r})+{\bf A}^{S}({\bf r})=
\sum_{i=0}^{N_T-1}(h{\bf j})_i\int\frac{d{\bf S}_i}{|{\bf r-r}_i|}
,\ee
where the superscripts $I$ and $S$ are designating the magnetic vector potential 
due to eddy currents and to the sink/source currents, respectively. The summation is over the $N_T$ FE triangles, while  the surface integral  is taken over the considered FE triangles  analytically.

The equation for the stream function $I$ is given by \cite{CVA2013, HOL2012} 
\be
\na\cdot(\frac{1}{d_w\sigma}\na I)
=
\R{\pa B_\perp}{\pa t}
=
\R{\pa (B^{pl}_\perp+B^{coil}_\perp+B^{I}_\perp+B^S_\perp+B^{Fe}_\perp)}{\pa t}
\hspace{3.cm} 
\lbq{eq:I}
\ee
where
 $B^{pl, coil, I, S, Fe}_{\perp}$ are representing different perpendicular to the wall magnetic field  components (due to plasma, coils, eddy currents, sink/source currents and iron core transformer).
 
  To close the system of equations, the Biot-Savart relation for $B$ is necessary.
  
\section{Energy principles for the wall currents} 
The tokamak wall configurations being complex  and presenting  different material interfaces,
solving the strong form (PDE) is not always efficient. Therefore, we have used a weak formulation - a finite
element method formulation - for our problem.
Thus,
$\gj^S$  was obtained by
minimizing the functional $W^S$ \cite{LEZ2015, LEZ20171, CVA20171, CVA20172, CVA20173}
 \be
W^S
=
\int\left\{
\underbrace{
\R{d_w\sigma(\na\gj^S)^2}{2}-j_\perp\gj^S}_{minim.\;gives\;Eq. (2)}
\right\}dS
-\oint\underbrace{\gj^Sd_w\sigma[(\wsn\times\na\gj^S)}_{S.C.\;\perp to\;the\;edges}\cdot d\vslc].
\lbq{eq:Wgk}
\ee
where  $\int dS$ is taken along the wall surface, 
 $\oint d\vslc$ is taken along the edges of the
conducting surfaces with the integrand representing the surface current
normal to the edges. This last integral takes into account the external
voltage applied to the wall edges  and vanishes as happens in typical cases.

For the divergence-free part of the surface current ${\bf i}$, the energy principle
looks like \cite{LEZ2015, CVA20171, CVA20172}
\be
\begin{aligned}
W^I\equiv
\R{1}{2}\int\Bigg\{\underbrace{
	\R{\pa(\wsi\cdot\waA^{I})}{\pa t}}_{inductive \;term\;due\; to\; {\bf i}}
+
\underbrace{
	\frac{1}{d_w\sigma}|\na I|^2}_{\;\;resistive\; loses}\\ 
+
\underbrace{
	2\left(\wsi\cdot\R{\pa\waA^{ext}}{\pa t}\right)}_{excitation\;by\;other\;sources}
\Bigg\}dS - 
\underbrace{
	\oint(\gj^E-\gj^S)\R{\pa I}{\pa \mathcal{\ell}}d\mathcal{\ell}}_{S.C.\;\perp\;to\;edges}.
\lbq{eq:WI1}
\end{aligned}
\ee

\section{Simulation of Eddy and Source/Sink currents}
\subsection{Numerical solution}
We have adopted a triangle electromagnetic representation of the thin wall,
based on the expressions for  $\waA$ and $\waB$ of a uniform 
surface current $\wsj$=const inside each triangle. The 
two energy functionals  for $\gj^S$  and  for $I$  constitute our electromagnetic wall model for the wall touching kink and vertical
modes.  The substitution of $I,\gj^S$ as a set of plane functions inside
triangles  leads to the finite
	element representation of $W^I,W^S$ as quadratic forms for unknowns
	$I,\gj^S$ in each vertex. The minimization of quadratic forms  $W^S$ and  $W^I$ 
	\bee
	\pa W^S/\pa \vec{\phi}^S=0,\;\; \pa W^I/\pa \vec{I}^n=0,\;\pa W^I/\pa \vec{\phi}^S=0,
	\eee
	leads to a linear systems of equations with Hermitian symmetric-positive definite matrices
		which can be solved using the Cholesky decomposition.
 
	The matrix equations obtained after minimization and decoupling between equations for eddy and SS currents (easily achieved by matrix multiplications) are \cite{CVA20171, CVA20172}
\be
	\vec{\phi}^S &=&-\left ({\bf W}^{SS}\right) ^{-1}\cdot \underbrace{\vec{j}_{\perp}}_{input}
	\nonumber \\
		\vec{I}^n&=&\underbrace{\vec{I}^{n-1}}_{input}-\widehat{{\bf R}}\cdot \underbrace{\vec{I}^{n-1}
		\Delta t}_{input}+\widehat{{\bf W}}^{IS}\cdot 
	\underbrace{\frac{\partial \vec{j}_{\perp}}{\partial t}\Delta t}_{input}
	- 
	\widehat{{\bf A}}^{IS}\cdot \underbrace{\frac{\partial (\vec{A}^{pl}+\vec{A}^{ext})}{\partial t}\Delta t}_{input}.
	\ee	
	with vector sources $\vec{j}_{\perp},\;\;\vec{A}^{pl},\;\;\vec{A}^{ext}$:
	\be
	&&\vec{j}_{\perp}\equiv\{j_{\perp,0},\;j_{\perp,1},\;j_{\perp,2},\;...,\;j_{\perp,N_V-1}\},\nonumber \\
	&&\vec{A}^{pl,ext}\equiv\{\vec{A}^{pl,ext}_0,\;\vec{A}^{pl,ext}_1,\;\vec{A}^{pl,ext}_2,\;...,\;
	\vec{A}^{pl,ext}_{N_V-1}\},
	\ee
	$N_v$ is the vertex number, $\Delta t$ is the ''wall-time-step'', while
the	superscript $n$ represents the time slice. Finally, the calculation of the wall currents is reduced to the  two relations Eqs. (8) implemented in our code.
	 
	In Table 1, the  size of the matrices for the 21744 triangles and 11223 vertexes of the FE discretization of ITER wall is given.
As output, the code returns the values of $\phi^S_i$ and $I_i$ in all vertexes, allowing the calculation 
	of the {\bf A} and {\bf B} of the wall currents in any point $\vec{r}$.
\begin{table}
\caption{\label{1} Matrices size for the 21744 triangles and 11223 vertexes of the FE discretization of ITER wall.}
\begin{center}
\begin{tabular}{ll}
\br
Matrix& Memory size [KB]\\
\mr
$({\bf W}^{SS})^{-1}$ & 984,030\\
	$\widehat{{\bf R}}$ & 855,106\\
	$\widehat{{\bf W}}^{SS} $& 917,305\\
	$\widehat{{\bf A}}^{IV}$ & 917,305\\
\br
\end{tabular}
\end{center}
\end{table}
In Figure 1 a finite element discretization of ITER wall is presented. 

\begin{figure}
\begin{center}
\hspace*{-.6cm}
\includegraphics[scale=.5]{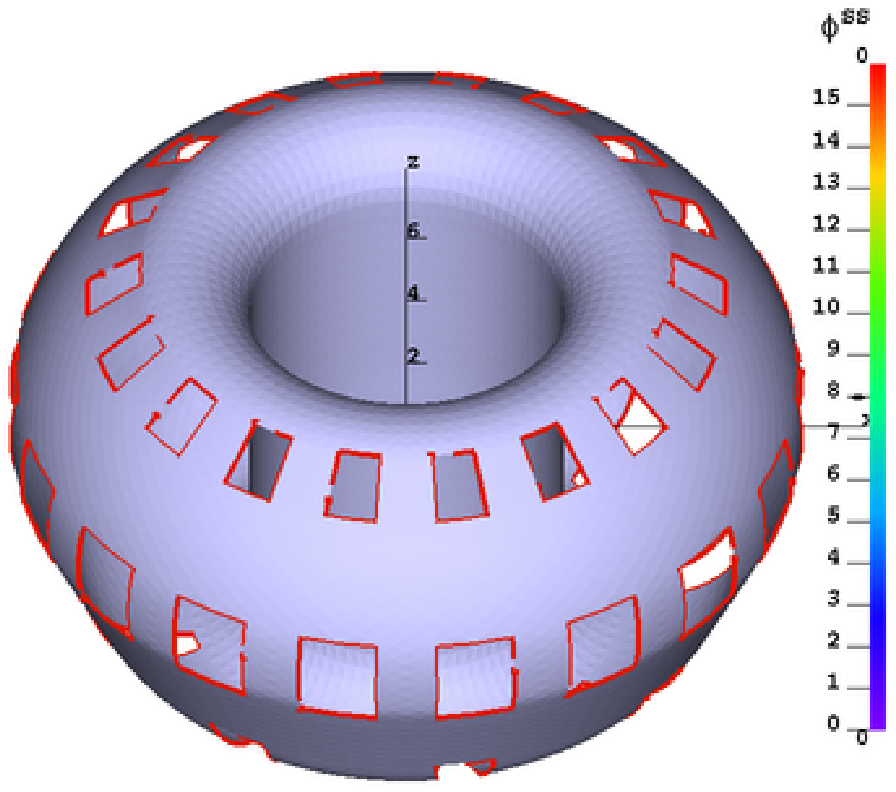}
\vspace*{1.8cm}
\hspace*{2.cm}
\includegraphics[scale=.6]{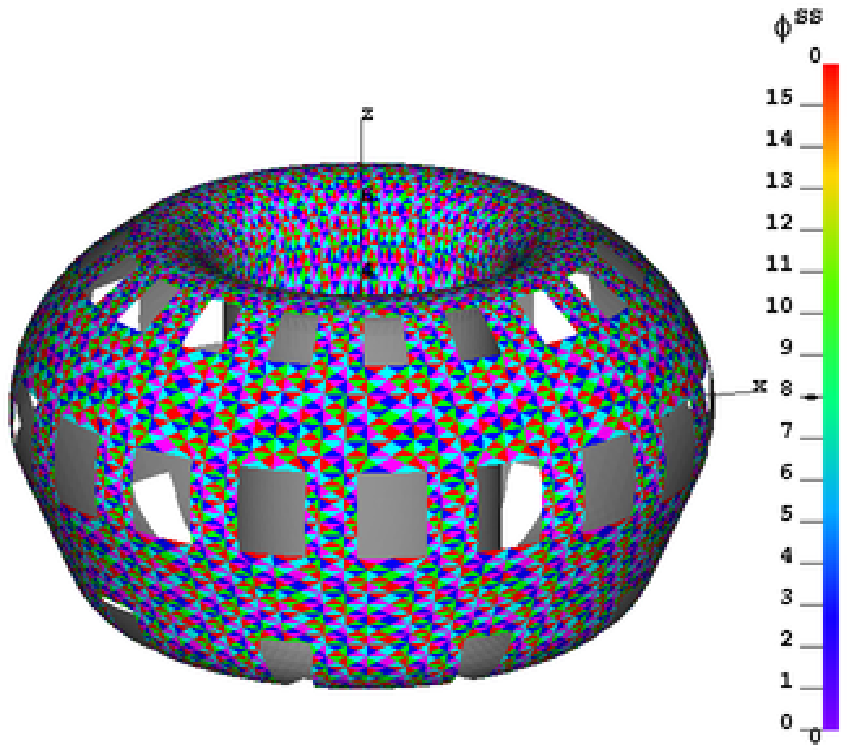}
\end{center}
\vspace*{-1.8cm}
\caption{\label{1} Finite edge elements (left) and full triangle finite elements  (21744) distribution in ITER wall (right).}
\end{figure}

 In Figure 2, an example of our calculation, with a conceived $j_{\perp}$ distribution is given.
 In our calculations, both wall thickness $d_w$ and conductivity $\sigma$ can be considered as variable too.
 For a ITER wall with 11223 vertexes and 21744 triangles, the generation of the matrix
and its Choleski decompositions takes $\sim 15'$, after this, the solution of the equation takes
several seconds for given $j_{\perp}$ distributions.

\begin{figure}
\begin{center}
\hspace*{-.6cm}
\includegraphics[scale=.5]{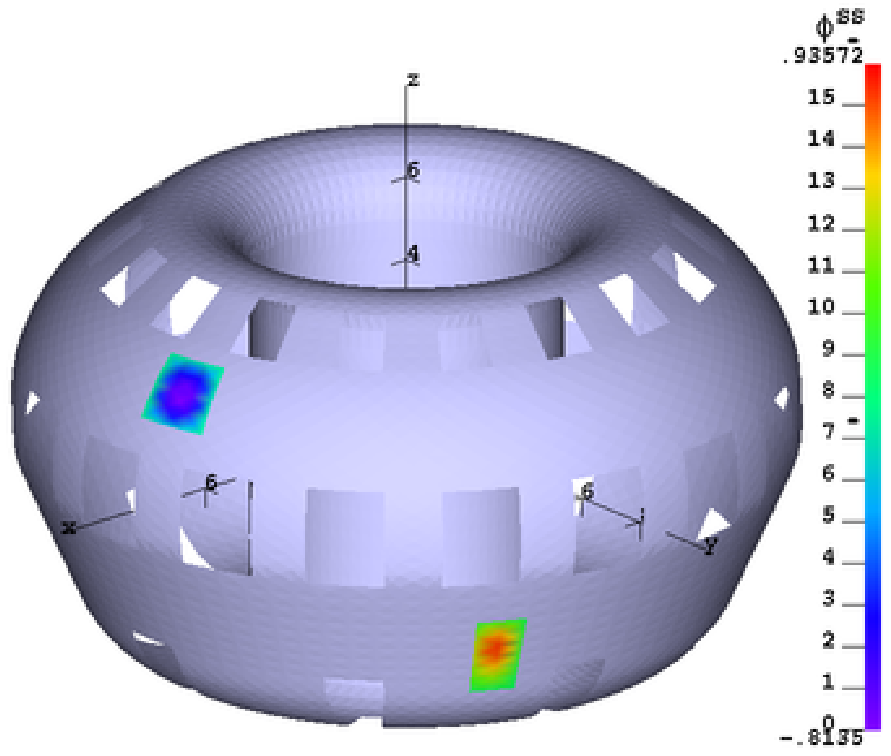}
\vspace*{1.5cm}
\hspace*{1.1cm}
\includegraphics[scale=.6]{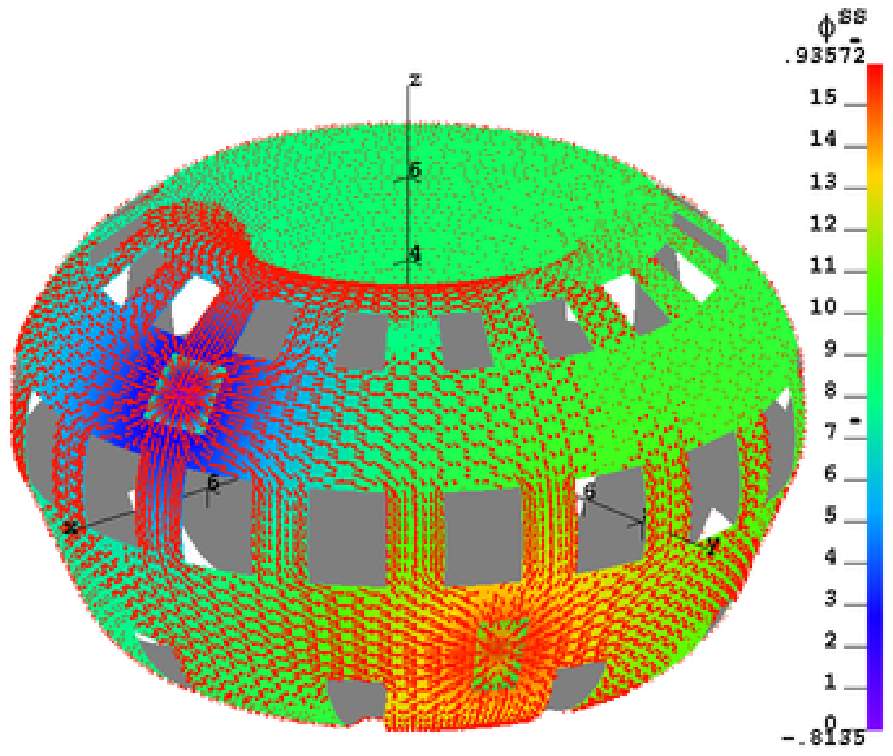}
\end{center}
\vspace*{-1.5cm}
\caption{\label{1} (left) Example of a conceived $j_{\perp}$ distribution in two locations: from the plasma to the 
wall and from the wall back to the plasma. (right) The correspondent solutions.
The generation of the matrix and its Choleski decompositions takes 
$\sim 15'$, after this,
the solution of the equation takes several seconds for given $j_{\perp}$ distributions.}
\end{figure}
\subsection{Analytical solution for $\phi^S$}
For a shell with elliptical cross-section and three holes with
the correspondent geometry in a curvilinear coordinate system $(u, v)$ in Fig. 3 (left) and considering 
 $d_w\sigma$=1, we have to solve the equation
\be
\nabla^2\phi^S=j_{\perp}(u,v),\;\; u={\rm toroidal\;coordinate},\;\;v={\rm poloidal\; coordinate,}
\ee
with pure homogeneous Neumann B.C. and the following existence condition which is satisfied:
\be
\begin{aligned}
\int_{\Omega}j_{\perp}d\Omega=\int_{\partial \Omega}\nabla \phi^S\cdot {\bf n}dS   \;\;\;\;\;\;
\;\;\;\;\;\;\;\;\;\;\;\;\;\;\;\;\;\;\;\;\;\;\;\;\;\;\;\;\;\;\;\;\;\;\;\;\;\;\;\;\;\;\;\;\;\;\;  \\ 
\Omega=\underbrace{\Omega_e}_{wall\;domain} \setminus 
\underbrace{\Omega_i}_{hole\;domain}\;\; \partial \Omega=\underbrace{\Gamma_e}_
{wall\;boundary} \cup \underbrace{\Gamma_i}_{hole\;boundary}.
\end{aligned}
\ee

The analytical $\phi(u,v)$ has been chosen in the form  [11, 12]
\be
\begin{aligned}
\phi^S(u,v)=\int G_u(u)du\cdot \int G_v(v)dv,\;\;\rm {with}
\;\;\;\;\;\;\;\;\;\;\;\;\;\;\;\;\;\;\;\;\;\;\;\;\;\;\;\;\;\;\;\;\;\;\;\\
G_u(u)=\Pi(u-u_{ik});\;\;G_v(v)=\Pi(v-v_{ik});\;\;i=0,...,3,\;k=1,2,
\end{aligned}
\ee
\begin{figure}
	\begin{center}
		\hspace*{-1.4cm}
	\includegraphics[scale=.7]{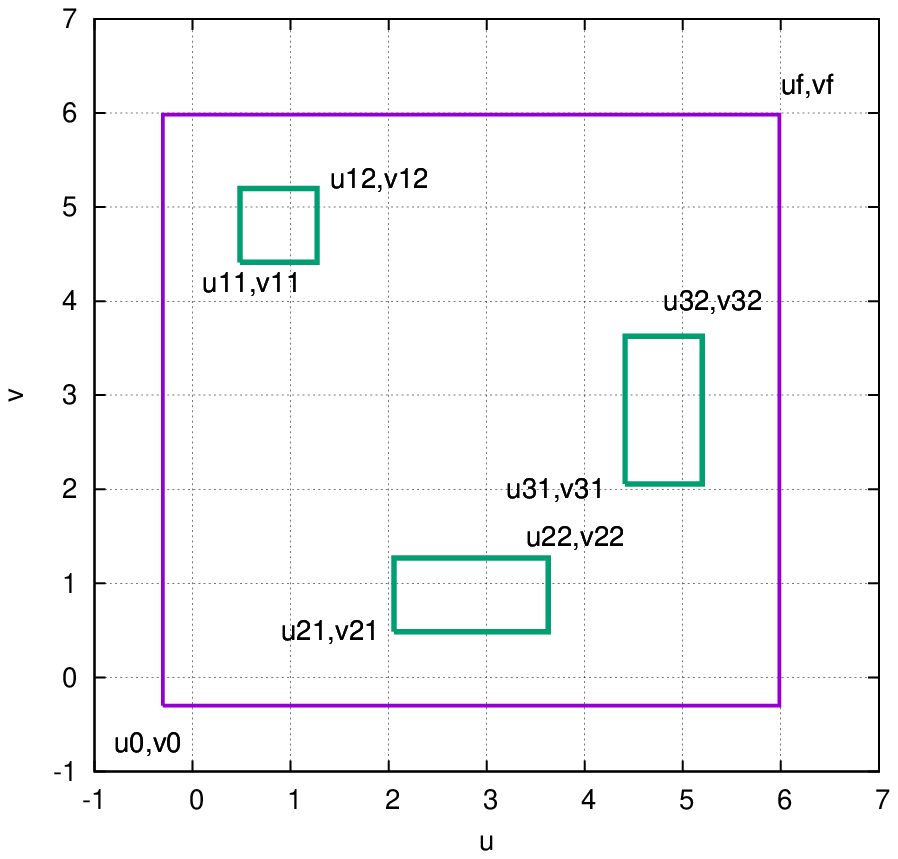}
	\vspace*{1.5cm}
\hspace*{1.1cm}
\includegraphics[scale=.7]{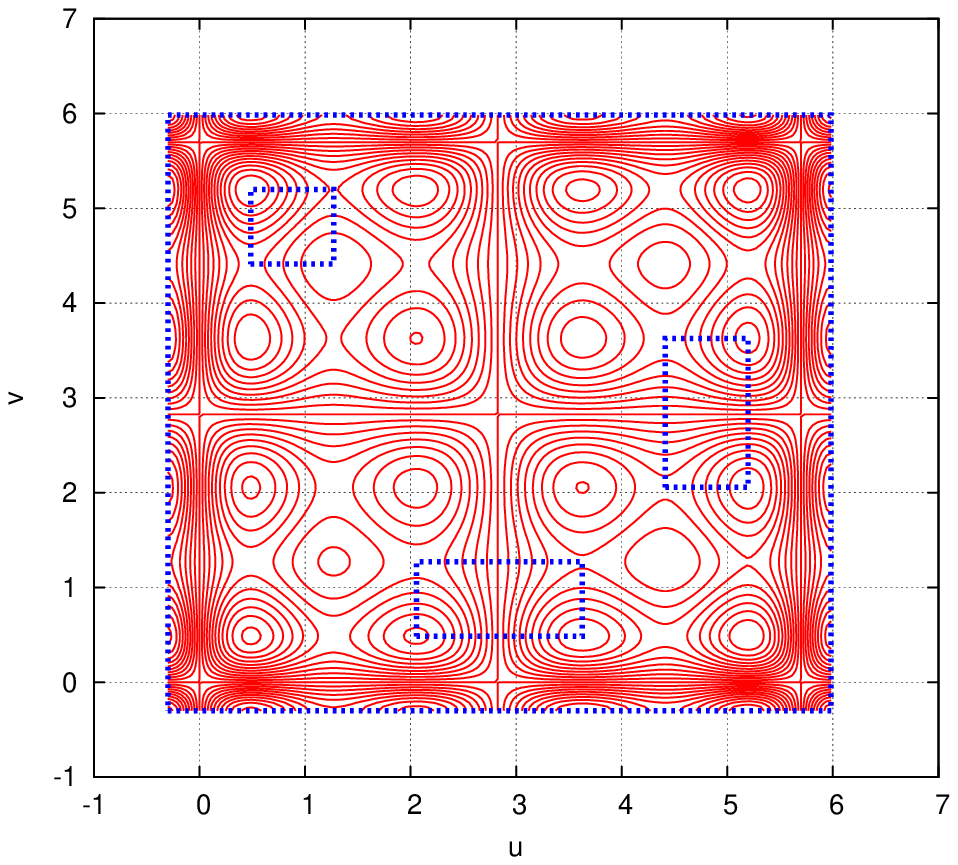}
	\end{center}
	\vspace*{-1.6cm}
	\caption{\label{1}(left) Tokamak wall with elliptical cross-section and three holes (in blue) - multiply connected test domain 
		$D(u,v)$ between the four rectangles in a 
		curvilinear coordinate system $(u,v)$; (right ) Distribution of the analytical $\phi^S(u,v)$ function. }
\end{figure}
The analytical solution is presented in Figure 3. If for one hole the relative error was of 0.003 for a grid with a mesh $32\times 32\times 4$, for three holes the error is $\approx$ 5 times greater.  This is due to the presence of many re-entry corners (L-shaped domains).   This problem has been solved by using conformal transformations so that the relative error became significantly smaller.
\subsection{Analytic solution for $I$}
Writing Eq. (5) in the same curvilinear coordinate system $(u,v,w)$, where two of the covariant basis 
vectors ${\bf r}_u\equiv \partial {\bf r}/{\partial u}$ and  
${\bf r}_v\equiv \partial {\bf r}/{\partial v}$ are tangential to the wall surface ($d_w\sigma$ has been
considered constant) 
\be
\frac{1}{D}\bigg \{\frac{\partial}{\partial u}
\bigg [ 
\bigg (\frac{g_{vv}}{D}\frac{\partial I}{\partial u}
-\frac{g_{uv}}{D}\frac{\partial I}{\partial v}
\bigg )\bigg ]+
\frac{\partial}{\partial v}
\bigg [ 
\bigg (\frac{g_{uu}}{D}\frac{\partial I}{\partial v}
-\frac{g_{uv}}{D}\frac{\partial I}{\partial u}
\bigg )\bigg ]=\frac{\partial B_{\perp}}{\partial t},
\ee
where, as before, $I$ is the stream function of the divergence free component (eddy currents), $g_{uu},\; g_{uv}$ and $g_{vv}$ are the covariant metric coefficients and $D$ is the 2D Jacobian at the wall 
surface. ${\bf B}$ is the magnetic field on the wall surface. We have
\be
\nabla \times {\bf i}=-d_w\sigma\frac{\partial {\bf B}}{\partial t},
\ee
and by integrating this equation on the surface $S_{\gamma}$ delimited by $\gamma$ curve (wall and/or hole boundaries) we obtain
\be
\int_{S_{\gamma}} \nabla \times {\bf i}d{\bf s}=\oint_{\gamma}{\bf i}d{\bf l}=-d_w\sigma
\int_{S_{\gamma}}{\bf B}d{\bf s}=-d_w\sigma\frac{\partial \Phi_{S_{\gamma}}}{\partial t}.
\ee
$\Phi^S$ is the magnetic flux through the $S_{\gamma}$ surface. Thus, Eq. (15) gives the necessary boundary conditions (of Dirichlet type) for the parabolic equation (13).

\section{ Calculation the iron core transformer influence in simulation of the wall response  
during VDE in JET tokamak
}
Due to the high non-linear dependence of the MHD solutions on the iron permeability of the iron-core tokamak transformer, the complexity of equilibrium and stability calculations increases considerably. A boundary integral equations method to calculate the magnetostatic part of the MHD equations is presented in
the following and has been used first by us, in a simplified form, for equilibrium calculation for the T15 tokamak \cite{CVA1990}. Let us consider an equivalent 2D magnetic circuit (i.e. with rotational symmetry) of the JET tokamak with the meridian cross-section presented in Fig. 4.
It is known that the surface current density distribution along a curve $\Gamma$ separating two homogeneous media and of constant magnetic permeabilities ($\mu^{out}$ and $\mu^{int}$ in the outer and inner domain respectively of the $\Gamma$ curve), is described by a Fredholm integral equation of second kind 
\be
\frac{1}{2}\mu_0 i^{Fe}(l)=\frac{\mu^{out}(l)-\mu^{int}(l)}{\mu^{out}(l)+\mu^{int}(l)}
\left ( B^{ext}_{\tau}(l)+\int_{\Gamma}b_{\tau}(l,l')i(l')dl'\right ),
\ee
where: $b_{\tau}$ is the tangential magnetic field component at the $\Gamma$ curve  in $l$ produced by a unit surface current in $l'$, $B^{ext}_{\tau}(l)$ is the tangential magnetic field component at the $\Gamma$ curve produced by an external known source.

\begin{figure}
	\begin{center}
		\hspace*{-1.4cm}
		\includegraphics[scale=.8]{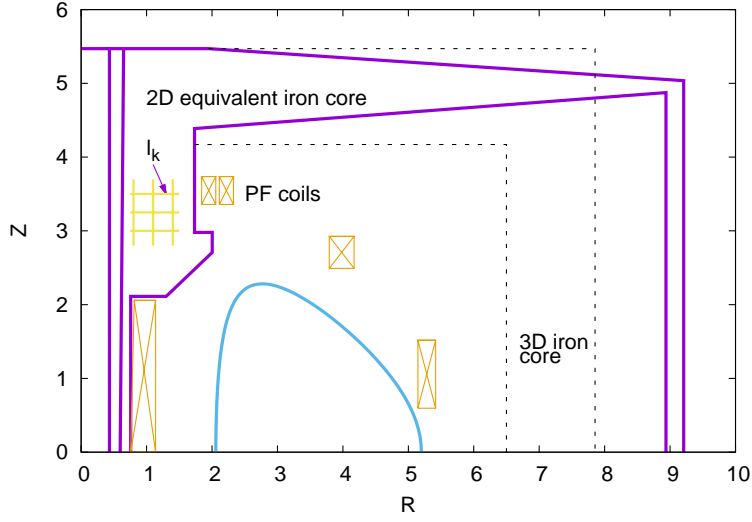}
	\end{center}
	\caption{\label{1} Meridian cross-section of the equivalent 2D magnetic circuit of the JET tokamak }
\end{figure}
For the real case, the $\Gamma$ curve was considered as the sum of all interfaces between subdomains of different permeabilities, then on each segment of curve (a Liapunov curve) the surface current density, supposed to be H\"older continuous and of bounded variation, admits a uniform convergent expansion
\be
i^{Fe}(l_k^n)=\sum_{i=0}^{p^n_k}C^n_{k,i}P_i(\lambda)w_i(\lambda),\;\;\lambda\in [-1,1],
\ee
where  now $\Gamma_k$ contours ($k=1,\;M$) with $n=1,\;N_k$ discrete segments for each contour have been considered. $C^n_{k,i}$ are the unknown coefficients, $P_i(\lambda)$ are orthogonal polynomials of 
order $p_k^n$, while $w_i(\lambda)$ are weight functions (given the azimuthally symmetric geometry, we have used Legendre polynomials, $w_i\equiv 1$). Thus, Eq. (16) becomes
\be
\begin{aligned}
\frac{\mu_0}{2i+1}A^n_{k,i} &=& \frac{\mu^{out}(l)-\mu^{int}(l)}{\mu^{out}(l)+\mu^{int}(l)}
\Bigg ( \int_{-1}^1B^{ext}_{\tau}(l_k^n)P_i(\lambda)dl  
\;\;\;\; \;\;\;\; \;\;\;\; \;\;\;\; \;\;\;\; \;\;\;\; \;\;\;\; \;\;\;\; \;\;\;\; \\ 
 &+&\sum_{j=1}^M\sum_{m=1}^{N_j}\sum_{l=0}^{p_j}A_{j,l}^m\Delta l^m_j\int_{-1}^1
 \int_{-1}^1b_{\tau}(l_k^n,\;l_j^m)P_i(\lambda)\;P_j(\nu)d\lambda d\nu\Bigg )
 \end{aligned}
\ee
Once the iron surface currents $i^{Fe}(l_k^n)$ have been determined, their contribution to the external magnetic field can be calculated. Evidently, an iterative feedback approach has to be considered.
\section{Summary}
Within the framework of a thin wall limit and a triangular representation of the wall surface, 
both divergence-free eddy and source/sink currents are represented by the same model of a uniform current density inside each triangle. This model is implemented in the SSC and the shell simulation code SHL. 
On request, our code received the status of {\it open source license}: to be used now by the entire EUROfusion community in modelling Wall Touching Kink Modes and  
Vertical Displacement Events.  Recently, our approach has been implemented successfully   into the JOREK-STARWALL code \cite{ART2018}.

 As a next step, in order to model a real disruption, we have to introduce the following input data in  our code 
\be
\vec{A}^{pl}+\vec{A}^{ext}=f_A(t,{\bf r}),\;\;
\vec{B}^{pl}+\vec{B}^{ext}=f_B(t,{\bf r}),\;\;
\vec{J}_{\perp}=f_J(t,{\bf r}).
\ee
\\ 
{\bf Acknowledgments}\\ 
This work has been  
partially carried out within the framework of the EUROfusion Consortium and has received funding from the Euratom research and training programme 2014–2018 under grant agreement no. 633053. The views and opinions expressed herein do not necessarily reflect those of the European Commission. 

\end{document}